%% file: cs_merge.tex
\documentclass{llncs}
\usepackage{graphicx}
\usepackage{amsmath}
\usepackage[ruled,linesnumbered]{algorithm2e}
\usepackage[noend]{algpseudocode}
\SetAlFnt{\scriptsize}
\usepackage{hyperref}
\usepackage{tabularx}
\usepackage{tabu}
\usepackage{booktabs}
\usepackage{listings}
\usepackage{float}
\usepackage{subfig}
\usepackage{enumitem}

\begin{document}
\title{Hierarchical Characteristic Set Merging for Optimizing SPARQL Queries in Heterogeneous RDF}

\author{Marios Meimaris\inst{1} \and George Papastefanatos\inst{1}}

\institute{ATHENA Research Center\\
\email{[m.meimaris, gpapas]@imis.athena-innovation.gr}} 


\maketitle

\begin{abstract}
Characteristic sets (CS) organize RDF triples based on the set of properties characterizing their subject nodes. This concept is recently used in indexing techniques, as it can capture the implicit schema of RDF data. While most CS-based approaches yield significant improvements in space and query performance, they fail to perform well in the presence of schema heterogeneity, i.e., when the number of CSs becomes very large, resulting in a highly partitioned data organization. In this paper, we address this problem by introducing a novel technique, for merging CSs based on their hierarchical structure. Our technique employs a lattice to capture the hierarchical relationships between CSs, identifies dense CSs and merges dense CSs with their ancestors, thus reducing the size of the CSs as well as the links between them. We implemented our algorithm on top of a relational backbone, where each merged CS is stored in a relational table, and we performed an extensive  experimental study to evaluate the performance and impact of merging to the storage and querying of RDF datasets, indicating significant improvements.

\end{abstract}

\input{section1_introduction}
\input{section2_relatedwork}
\input{section3_algorithm}
\input{section4_evaluation}

\input{section5_conclusions}
\bibliographystyle{abbrv}
\bibliography{sigproc}
\end{document}

%% file: section1_introduction.tex
\section{Introduction}
Recent works in the state of the art in RDF data management have shown that extraction and exploitation of the implicit schema of the data can be beneficial in both storage and SPARQL query performance \cite{pham2015deriving}\cite{pham2016exploiting}\cite{meimaris2017extended}\cite{montoya2017odyssey}.  In order to organize on disk, index and query triples efficiently, these trends heavily rely on two structural components of an RDF dataset, namely (i) the notion of \textit{characteristic sets }(CS), i.e., different property sets that characterize subject nodes, and (ii) the join links between CSs. For the latter, in our previous work, we introduced \textit{Extended Characteristic Sets} (ECS)\cite{meimaris2017extended}, which are typed links between CSs that exist only when there are object-subject joins between their triples, and we showed how RDF data management can rely extensively on CSs and ECSs for both storage and indexing, yielding significant performance benefits in heavy SPARQL workloads. However, this approach failed to address schema heterogeneity in loosely-structured datasets, as this implied a large number of CSs and ECSs (e.g., Geonames contains 851 CSs and 12136 CS links), and thus, skewed data distributions that impose large overheads in the extraction, storage and disk-based retrieval\cite{pham2015deriving}\cite{meimaris2017extended}.

In this paper, we exploit the hierarchical relationships between CSs, as captured by subsumption of their respective property sets, in order to merge related CSs. We follow a relational implementation approach by storing all triples corresponding to a set of merged CSs into a separate relational table and by executing queries through a SPARQL to SQL transformation. Although, alternative storage technologies can be considered (key-value, graph stores,etc), we have selected well-established technologies and database systems for the implementation of our approach, in order to take advantage of existing data indexing and query processing techniques that have been proven to scale efficiently in large and complex datasets. To this end, we present a novel system, named \textit{raxonDB}, that exploits these hierarchies in order to merge together hierarchically related CSs and decrease the number of CSs and the links between them, resulting in a more compact schema with better data distribution. The resulting system, built on top of PostgreSQL, provides significant performance improvements in both storage and query performance of RDF data. 

In short, our contributions are as follows: 
\begin{itemize}
\item We introduce a novel CS merging algorithm that takes advantage of CS hierarchies,
\item we implement \textit{raxonDB}, an RDF engine built on top of a relational backbone that takes advantage of this merging for both storing and query processing,
\item we perform an experimental evaluation that indicates significant performance improvements for various parameter configurations.
\end{itemize}

%% file: section2_relatedwork.tex
\section{Related Work}
RDF data management systems generally follow three storage schemes, namely \textit{triples tables}, \textit{property tables}, and \textit{vertical partitioning}. A triples table has three columns, representing the subject, predicate and object (SPO) of an RDF triple. This technique replicates data in different orderings in order to facilitate sort-merge joins. RDF-3X \cite{neumann2010RDF} and Hexastore \cite{weiss2008hexastore} build tables on all six permutations of SPO. Built on a relational backbone, Virtuoso \cite{erling2010virtuoso} uses a 4-column table for quads, and a combination of full and partial indexes. These methods work well for queries with small numbers of joins, however, they degrade with increasing sizes, unbound variables and joins.

\textit{Property Tables} places data in tables with columns corresponding to properties of the dataset, where each table identifies a specific resource type. Each row identifies a subject node and holds the value of each property. This technique has been implemented experimentally in Jena \cite{wilkinson2006jena} and DB2RDF \cite{bornea2013building}, and shows promising results when resource types and their properties are well-defined. However, this causes extra space overhead for null values in cases of sparse properties \cite{abadi2007scalable}. Also, it raises performance issues when handling complex queries with many joins, as the amounts of intermediate results increase \cite{janik2005brahms}. 

Vertical partitioning segments data in two-column tables. Each table corresponds to a property, and each row to a subject node \cite{abadi2007scalable}. This provides great performance for queries with bound objects, but suffers when the table sizes have large variations in size \cite{sidirourgos2008column}. TripleBit \cite{yuan2013triplebit} broadly falls under vertical partitioning. In TripleBit, the data is vertically partitioned in chunks per predicate. While this reduces replication, it suffers from the same problems as property tables. It does not consider the inherent schema of the triples in order to speed up the evaluation of complex query patterns.

In distributed settings, a growing body of literature exists, with systems such as Sempala \cite{schatzle2014sempala}, H2RDF \cite{papailiou2014h} and S2RDF \cite{schatzle2016s2rdf}. However, these are based on parallelization of centralized indexing and query evaluation schemes.

For these reasons, latest state of the art approaches rely on implicit schema detection in order to derive a hidden schema from RDF data and index/store triples based on this schema. Furthremore, due to the tabular structure that tends to implicitly underly RDF data, recent works have been implemented in relational backbones. In our previous work \cite{meimaris2017extended}, we defined \textit{Extended Characteristic Sets (ECSs)} as typed links betwen CSs, and we showed how ECSs can be used to index triples and greatly improve query performance. In \cite{pham2015deriving}, the authors identify and merge CSs, similar to our approach, into what they call an \textit{emergent schema}. However, their main focus is to extract a human-readable schema with appropriate relation labelling. They do not use hierarchical information of CSs, rather they use semantics to drive the merging process. In \cite{pham2016exploiting} it is shown how this \textit{emergent schema} approach can assist query performance, however, the approach is limited by the constraints of human-readable schema discovery. In our work, query performance, indexing and storage optimization are the main aims of the merging process, and thus we are not concerned about providing human-readable schema information or any form of schema exploration. In \cite{montoya2017odyssey}, the authors use CSs and ECSs in order to assist cost estimation for federated queries, while in \cite{gubichev2014exploiting}, the authors use CSs in order to provide better triple reordering plans. To the best of our knowledge, this is the first work to exploit hierarchical CS relations in order to merge CSs and improve query performance.

%% file: section3_algorithm.tex
\section{Hierarchical CS Merging}

\subsection{Preliminaries}
The RDF model does not generally enforce structural rules in the representation of triples; within the same dataset there can be largely diverse sets of predicates emitting from nodes of the same semantic type \cite{meimaris2017extended,pham2015deriving,neumann2011characteristic}. \textit{Characteristic Sets} (CS)\cite{neumann2011characteristic} capture this diversity by representing implied node types based on the set of properties they emit. Formally, given a collection of triples $D$, and a node $s$, the characteristic set $cs(s)$ of $s$ is $cs(s) = \{p \mid \exists o: (s, p, o) \in D\}$.

The set of properties of a CS $cs_i$ is denoted with $P_i$. Furthermore, in a given dataset, each CS represents a set of records identified by a subject node, and all of the values of the subject node (i.e., objects) for the predicates in $P_i$. We denote the set of all records of $cs_i$ as $r_i$, while $cs_i$ is represented by a relational table $c_i$ that is defined by these two elements, i.e., $c_i = (P_i, r_i)$. The tuples in $c_i$ are of the form $(s,p_{i,1},\dots, p_{i,k})$, where $s$ is the identifier column (e.g. URI) of a subject node and $p_{i,1}, p_{i,2}, \dots ,p_{i,k}$ are the values, i.e. object nodes, of the properties in $P_i$ for $s$. In the context of this paper, with the term \textit{Characteristic Set} we will refer collectively to the properties and records of a CS, i.e., its relational table, rather than just the set of properties proposed in the original definition, for the sake of simplicity. 

Within a given dataset, CSs often exhibit hierarchical relationships, as a result of the overlaps in their comprising sets of properties. For example, consider two CSs, $c_1, c_2$, describing human beings, with $P_{1} = \{type, name\}$ and $P_{2} = \{type, name, marriedTo\}$. It can be seen that $P_{1} \subset P_{2}$ and thus $c_1$ is a parent of $c_2$. This relationship entails an overlap of properties that define the CSs, and can be exploited in order to provide a means to merge common CSs based on the specialization or generalization of the node types they describe. In what follows, we formally define the notions of CS \textit{subsumption}, \textit{hierarchy} and \textit{ancestral sub-graphs}. 

\begin{figure}[t]
\centering
\includegraphics[width=1\textwidth]{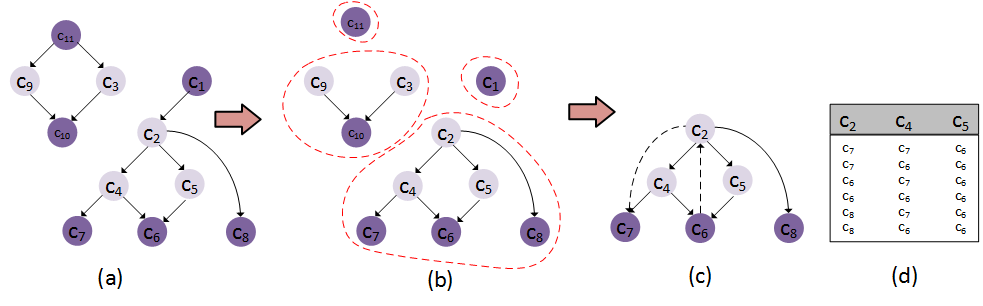}
\caption{(a) A CS hierarchy graph with dense nodes colored in deep purple, (b) the connected components derived by cutting off descendants from dense nodes, (c) a connected component with dashed lines representing inferred hierarchical relationships, (d) all possible assignments of dense nodes to non-dense nodes.}\label{fig:combinations}
\end{figure}

\textit{Definition 1}. \textbf{\textit{(CS Subsumption)}}.
 Given two CSs, $c_i$ and $c_j$, and their property sets $P_i$ and $P_j$, then $c_i$ subsumes $c_j$, or $c_i \succ c_j$, when the property set of $c_i$ is a proper subset of the property set of $c_j$, or $P_i \subset P_j$. This subsumption forms parent-child relationships between CSs. CS subsumption relationships can be seen in Figure \ref{fig:combinations}(a) as directed edges between nodes. The set of all parent-child relationships defines a CS hierarchy as defined in the following. 

\textit{Definition 2}. \textbf{\textit{(CS Hierarchy and Inferred Hierarchy)}}.
 CS subsumption creates a partial ordering that essentially defines a \textit{hierarchy} such that when $c_i \succ c_{j}$, then $c_{i}$ is a parent of $c_{j}$. Formally, a CS hierarchy is a graph lattice $L = (V, E)$ where $V \in C$ and $E \in (V \times V)$. A directed edge between two CS nodes $c_1, c_2$ exists in $L$, when $c_1 \succ c_2 $ and there exists no other $c_i$ such that $c_1 \succ c_i \succ c_2$. An example CS hierarchy can be seen in Figure \ref{fig:combinations}(a). Given a hierarchy $L$, we denote the \textit{hierarchical closure} of $L$ with $L_{c}$, so that $L_{c}$ extends $L$ to contain inferred edges between hierarchically related nodes that are not consecutive, e.g. a node and its grandchildren. An example inferred hierarchy can be seen in Figure \ref{fig:combinations}(c) for a sub-graph of the graph in Figure \ref{fig:combinations}(a), with the inferred relationships in dashed lines. In the remainder of this paper, we refer to $L_{c}$ as the \textit{inferred hierarchy} of $L$.

\textit{Definition 3}. \textbf{\textit{(CS Ancestral Sub-graphs)}}.
 Given an inferred hierarchy $L_{c} = (V, E)$, a CS $c_{base}$ and set of CSs $c_1, \dots, c_k$, then $a = (V^{'}, E^{'})$ is an ancestral sub-graph with $c_{base}$ as the lowermost child when $\forall i \in [1..k]$, it holds that $c_{i} \succ c_{base}$, and $(c_{i}, c_{base}) \in E^{'}$. This means that any sub-graph with $c_{base}$ as a sink node will be an ancestral sub-graph of $c_{base}$. Thus, it holds that $a \subset L_c$. For instance, in Figure \ref{fig:combinations}(c), nodes $c_7, c_4, c_2$ form an ancestral sub-graph with $c_7$ as the base CS. Similarly, nodes $c_6, c_4, c_2$ and $c_6, c_5, c_2$ form ancestral sub-graphs with $c_6$ as base CS. 
 
Logically, we map each CS to a relational table, so that for a CS $c_i$ we create a relational table $t_i = (s, p_{i,1}, p_{i,2}, \dots, p_{i,k})$, where $s$ is the id of the subject and $p_{i,1}\dots, p_{i,k}$ are the properties that belong to $P_i$, and then we use the CS hierarchy in order to merge the nodes of an ancestral sub-graph with $c_i$ as base into a single table. Specifically, we exploit the property set overlap in order to merge together smaller parent CSs with larger child CSs, in order to minimize the effect of NULL values that will appear for properties in smaller CSs that do no exist in the larger CSs. Thus, $c_{base}$ will be the most specialized CS in its ancestral sub-graph. For this reason, we define a merge operator, $hier\_merge$, as follows.

\begin{figure}[t]
\centering
\includegraphics[width=0.4\textwidth]{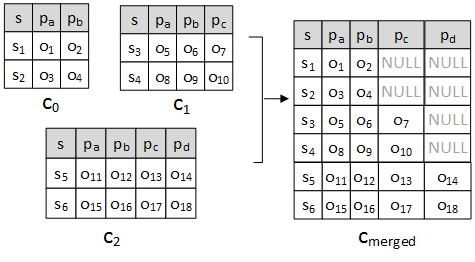}
\caption{Merging the tables of $c_0$, $c_1$ and $c_2$.}\label{fig:table_merging}
\end{figure}

\textit{Definition 4}. \textbf{\textit{(Hierarchical CS Merge)}}.
 Given an ancestral sub-graph $a = (V^{'}, E^{'})$, where $V^{'} = \{c_1 = (P_1, r_1), c_2 = (P_2, r_2),\dots, c_k=(P_k, r_k)\}$ as defined above, then a hierarchical merge of $a$ is given as follows: $hier\_merge(a) = c_{a}$, where $c_a = (P_1, r_a)$. Here, $P_1$ is the most specialized property set in $a$, as $c_1$ does not have any children in $a$, while $r_a =  \bigcup\limits_{i=1}^{k}r_i^{'}$ is the UNION of the records of all CSs in $V^{'}$, where $r_i^{'}$ is the projection of $r_i$ on $P_1$. This means that $r_i^{'}$ will contain NULL values for all the non-shared properties of $P_1$ and $P_i$, i.e., $P_1 \setminus P_i$. In essence, $hier\_merge$ is an \textit{edge contraction} operator that merges all nodes of an ancestral sub-graph into one, while removing the edges that connect them. For instance, assume that $V^{'} = \{c_0 = (P_0, r_0), c_1 = (P_1, r_1), c_2 = (P_2, r_2)\}$ is the set of vertices of an ancestral sub-graph with three CSs, with $P_0 = \{p_a,p_b\}$, $P_1 = \{p_a, p_b, p_c\}$ and $P_2 = \{p_a, p_b, p_c, p_d\}$. Thus, $c_0 \succ c_1 \succ c_2$. Hierarchical merging can be seen in Figure \ref{fig:table_merging}. 

\textit{Definition 5}. \textbf{\textit{(Merge Graph)}}.
Given an inferred CS hierarchy $L_{c} = (V,E)$, a merge graph is a graph $L^{'}=(V^{'}, E^{'})$ that consists of a set of $n$ ancestral sub-graphs, and has the following properties: (i) $L^{'}$ contains all nodes in $L$ such that $V^{'} \equiv V$, i.e., it covers all CSs in the input dataset, (ii) $L^{'}$ contains a subset of the edges in $L$ such that $E^{'} \subset E$, (iii) each node is contained in exactly one ancestral sub-graph $a_i$, (iv) all ancestral sub-graphs are pair-wise disconnected, i.e., there exist no edges between the nodes of different ancestral sub-graphs. Thus, each ancestral sub-graph can be contracted into one node unambiguously, using the $hier\_merge$ operator. Also, the total number of relational tables will be equal to the number of ancestral sub-graphs in the merge graph.


\textbf{Problem Formulation. }Given an inferred CS hierarchy $L_{c} = (V, E)$, the problem is to find a merge graph $L^{'} = (V, E^{'})$ in the form of a set of disconnected ancestral sub-graphs, that provides an optimal way to merge CS nodes. In other words, the problem is to find the best set of ancestral sub-graphs from an inferred hierarchy $L_{c}$ that minimize an objective cost function $cost(x)$, or more formally:
\begin{align}
L^{'} = argmin_{x \subset L_c}cost(x)
\end{align}


This formulation entails several problems. First, the notion of cost depends on possibly subjective factors, such as the query workload, the storage technology, the input dataset and so on. There is no universal cost model that can be deployed in order to assess the effectiveness of a merge graph. Moreover, neither the number of ancestral sub-graphs, nor the set of sub-graph roots is known as part of the input. A CS hierarchy of $n$ nodes can potentially create $2^n$ sub-graphs, while the number of possible sub-graph roots is also exponential with the respect to the hierarchy size. Thus, given an arbitrary cost function, this is a problem of \textit{non-uniform graph partitioning} on the inferred hierarchy $L_c$, which is known to be NP-Hard. That is, even with a deployed cost model, it is still an exponential problem to enumerate all possible sub-graphs and find the one with the minimum cost. For these reasons, we approach the problem by deploying a set of rules and heuristics that find a good merge graph efficiently and offer improved storage and query performance, as will be shown in the experiments. 

\subsection{CS Retrieval and Merging}


The primary focus of this work is to improve the efficiency of the storage and query capabilities of relational RDF engines by exploiting the implicit schema of the data in the form of CSs. However, CS merging results in several problems that need to be addressed in this context. These are discussed in what follows.

First, the problem of selecting ancestral sub-graphs is a computationally hard one, as mentioned earlier. For this reason, we rely on a simple heuristic in order to seed the process and provide an initial set of ancestral sub-graph \textit{sink nodes}, that will form the bases of the final merged tables, as defined in Definition 3. For this, we identify \textit{dense} CS nodes in the hierarchy (i.e, with large cardinalities) and use these nodes as the bases of the ancestral sub-graphs. While node density can be defined in many different ways, in the context of this work we define a $c_i$ to be dense, if its cardinality is larger than a linear function of the maximum cardinality of CSs in $D$, i.e., a function $d: N \rightarrow R$, with $d(c_i) = m \times |r_{max}|$. Here, $m \in \left[0,1\right]$ is called the \textit{density factor}, and $r_{max}$ is the cardinality of the largest CS in $D$. This means that, by definition, if $m = 0$, no CSs will be merged, because all CSs will be considered dense and thus each CS will define its own ancestral sub-graph, while if $m = 1$, all no ancestral sub-graph will be defined, and all CSs will be merged to one large table, as no CS has a cardinality larger than that of the largest CS. With a given $m$, the problem is reduced to finding the optimal ancestral sub-graph for each given dense node. 

Second, merging tables results in the introduction of NULL values for the non-shared columns, which can degrade performance. Specifically, merging CSs with different attribute sets can result in large numbers of NULL values in the resulting table. Given a parent CS $c_1 = (P_1, r_1)$ and a child CS $c_2 = (P_2, r_2)$ with $\left\vert{P_1}\right\vert < \left\vert{P_2}\right\vert$ and $\left\vert{r_1}\right\vert >> \left\vert{r_2}\right\vert$, the resulting $\left\vert{P_2 \setminus P_1}\right\vert \times \left\vert{r_1}\right\vert$ NULL cells will be significantly large compared to the total number of $r_1+r_2$ records, thus potentially causing poor storage and querying performance\cite{pham2015deriving}. For this reason, CS merging must be performed in a way that will minimize the presence of NULL values. The following function captures the NULL-value effect of the merge of two CSs $c_i = (P_i, r_i), c_j = (P_j,r_j)$ with $c_i \succ c_j$:
\begin{align}
r_{null}(c_i,c_j) = \frac{|P_j \setminus P_i| \times |r_i|}{(|r_j|)}
\end{align}
Intuitively, $r_{null}$ represents the ratio of null values to the cardinality of the base CS in the merge. The numerator of the fraction represents the total number of cell values that will be null, as the product of the number of non-shared properties and the cardinality of the parent CS. The denominator represents the cardinality of the base CS. Hence, the base CS must be a descendant (i.e., with more properties) in order to minimize the presence of NULLs.

In order to assess an ancestral sub-graph, we use a generalized version of $r_{null}$ that captures the NULL value effect on the whole sub-graph:
\begin{align}
r^{g}_{null}(g)|_{c_{d}} = \frac{\sum_{i=1}^{|g|}{|P_{d} \setminus P_i| \times |r_i|}}{|r_d|+\sum_{i=1}^{|g|}(|r_i|)}
\end{align}
Here, $c_d = (P_d, r_d)$ is the dense root of sub-graph $g$. However, merging a parent to a dense child changes the structure of the input graph, as the cardinality of the dense node is increased. To accommodate this, we define a cost function that works on the graph level, as follows:

\begin{align}
cost(g) = \sum_{i=1}^{n}{r^{g}_{null}(g_i)|_{c_{di}}}
\end{align}
where $n$ is the number of dense nodes, $c_{di}$ is a dense node and $g_i$ is the ancestral sub-graph with $c_{di}$ as the base node.


Given this cost model and a pre-defined set of dense nodes, our exact algorithm will find the optimal sub-graph for each dense node. An inferred hierarchy graph can be converted to a set of connected components that are derived by removing the outgoing edges from dense nodes, since we are not interested in merging children to parents, but only parents to children. An example of this can be seen in Figure \ref{fig:combinations}(b). For each component, we can compute $cost(g)$ as the sum of the costs of these components. The main idea is to identify all connected components in the CS graph, iterate through these components, enumerate all sub-graphs within the components that start from the given set of dense nodes, and select the optimal partitioning for each component. 

The algorithm can be seen in Algorithm \ref{alg:optimalMerge}. The algorithm works by first identifying all connected components of the inferred hierarchy (Line 2). Identifying connected components is trivially done using standard DFS traversal, and is not shown in the Algorithm. Then, we iterate each component (Line 3), and for each component, we generate all possible sub-graphs. Then, we calculate the cost of each sub-graph (Line 7) and if it is smaller than the current minimum, the minimum cost and best sub-graph are updated (Lines 8-9). Finally, we add the best sub-graph to the final list (Line 11) and move to the next component. 

To generate the sub-graphs, we do not need to do an exhaustive generation of $2^n$ combinations, but we can rely on the observation that each non-dense node must be merged to exactly one dense node. Therefore, sub-graph generation is reduced to finding all possible assignments of dense nodes to the non-dense nodes. An example of this can be seen in figure \ref{fig:combinations}. In the figure, nodes $c_2, c_4, c_5$ are non-dense, while nodes $c_6, c_7, c_8$ are dense. All possible and meaningful sub-graphs are enumerated in the table at the right of the figure, where we assign a dense node to each of the non-dense nodes. An assignment is only possible if there exists a parent-child relationship between a non-dense node and a dense node, even if it is an inferred one (e.g. $c_2$ is an inferred parent of $c_7$). Hence, the problem of sub-graph generation becomes one of generating combinations from different lists, by selecting one element from each list. The number of lists is equal to the number of non-dense nodes, and the elements of each list are the dense nodes that are related to the non-dense node.

\textbf{Complexity Analysis. } Assuming that a connected component $g$ has $k$ non-dense nodes and $d$ dense nodes, and each non-dense node $k_i$ is related to $e(k_i)$ dense nodes, then the number of sub-graphs that need to be enumerated are $\prod_{i=1}^{k}{e(k_i)}$. In the example of figure \ref{fig:combinations}, the total number of sub-graphs is $e(c_2) \times e(c_4) \times e(c_5) = 3 \times 2 \times 1 = 6$. In the worst case all $k$ nodes are parents of all $d$ nodes. Then, the number of total sub-graphs is $k^d$, which makes the asymptotic time complexity of the algorithm $O(k^d)$.

\begin{algorithm}
 \KwData{An inferred hierarchy lattice $L_c$ as a adjacency list , and a set of dense CSs $D$}
 \KwResult{A set of optimal ancestral sub-graphs}
 	init $finalList$\;
  	$connectedComponents \gets findConnectedComponents(L_c)$\;
  	\For {\textbf{each} $connectedComponent$}{
  		init $min \gets MAX\_VALUE$\;
  		init $bestSubgraph$ \;
  		\While {$next \gets connectedComponent.generateNextSubgraph()$} {
  			\uIf{$cost(next) < min$}{
  				$min \gets cost(next)$\;
  				$bestSubgraph \gets next$\;
  			}
  		}
  		$finalList.add(bestSubgraph)$\;
	}
	return $finalList$\;
 \caption{\emph{optimalMerge}}\label{alg:optimalMerge}
\end{algorithm}




\subsection{Greedy Approximation}
For very small $d$ (e.g. $d < 4$), the asymptotic complexity of $O(k^d)$ is acceptable. However, in real-world cases, the number of connected components can be small, making $d$ large. For this reason, we introduce a heuristic algorithm for approximating the problem, that does not need to enumerate all possible combinations, but instead relies on a greedy objective function that attempts to find the local minimum with respect to our defined cost model for each non-dense node. Note that it lies beyond the scope of this work to compute the degree of approximation to the optimal solution, however, in our experiments, the heuristic solution is shown to provide significant performance gains. 

The main idea behind the algorithm is to iterate the non-dense nodes, and for each non-dense node, calculate the $r_{null}$ function and find the dense node that minimizes this function for the given non-dense node. Then, the cardinalities will be recomputed and the next non-dense node will be examined. The algorithm can be seen in Algorithm \ref{alg:greedyMerge}. In the beginning, the algorithm initiates a hash table, $mergeMap$, with an empty list for each dense node (Lines 1-4). Then, the algorithm iterates all non-dense nodes (Line 5), and for each dense node, it calculates the cost $r_{null}$ of merging it to each of its connected dense nodes (Lines 5-13), keeping the current minimum cost and dense node. In the end, the current non-dense node is added to the list of the dense node that minimizes $r_{null}$ (Line 14). Notice that we do not need to split the hierarchy into connected components in order for $greedyMerge$ to work. 

\textbf{Complexity Analysis. }Given $k$ non-dense nodes and $d$ dense nodes, where each non-dense node $k_i$ is related to $e(k_i)$ dense nodes, the $greedyMerge$ algorithm needs $\sum_{i=1}^{k}{e(k_i)}$ iterations, because we need to iterate all $e(k_i)$ nodes for each $k_i$. In the worst case, every $k_i$ is related to all $d$ dense nodes, requiring $kd$ iterations. Assuming a constant cost for the computation of $r_{null}$, then the asymptotic complexity of the greedy algorithm is $O(kd)$, which is a significant performance when compared to the exponential complexity of $optimalMerge$.

\begin{algorithm}
 \KwData{A hash table $p$ mapping non-dense CSs to their dense descendants, a set of dense CSs $D$, and a set of non-dense CSs $K$}
 \KwResult{A hash table mapping dense CSs to sets of non-dense CSs to be merged}
 	init $mergeMap$\;
 	\For {\textbf{each} $d \in D$}{ 	
 		$mergeMap.put(d, new List())$\;
 	}
  	\For {\textbf{each} $k \in K$}{
  		$min \gets MAX\_VALUE$\;
  		init $bestDense$\;
	  	\For {\textbf{each} $d_k \in p.get(k)$}{
	  		$cost \gets r_{null}(k, d_k)$\;
	  		\uIf{$cost < min$}{
	  			$min \gets cost$\;
	  			$bestDense \gets d_k$\;
	  		}
	  	}
	  	$mergeMap.get(bestDense).add(k)$\;
	}
	return $mergeMap$\;
 \caption{\emph{greedyMerge}}\label{alg:greedyMerge}
\end{algorithm}

Obviously, this process does not necessarily cover all CSs of the input dataset. The percentage of the dataset that is covered by this process is called \textit{dense CS coverage}. The remainder of the CSs that are not contained by any merge path are aggregated into one large table containing all of their predicates. If the total coverage of the merging process is large, then this large table does not impose a heavy overhead in query performance, as will be shown in the experiments. Finally, we load the data in the corresponding tables. 

\begin{figure}[ht]
\centering
\includegraphics[width=1\textwidth]{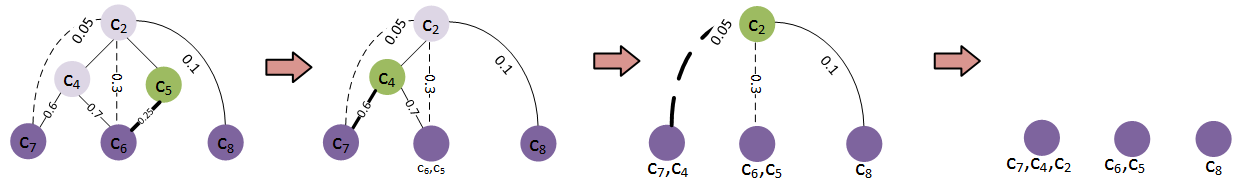}
\caption{An example of greedy merging. Dense nodes are coloured in deep purple. At each step, the non-dense node under examination is coloured with green, while the edge that minimizes $r_{null}$ can be seen in bold.}\label{fig:path_pruning}
\end{figure}

\subsection{Implementation}

\begin{figure}[t]
\centering
\includegraphics[scale=0.35]{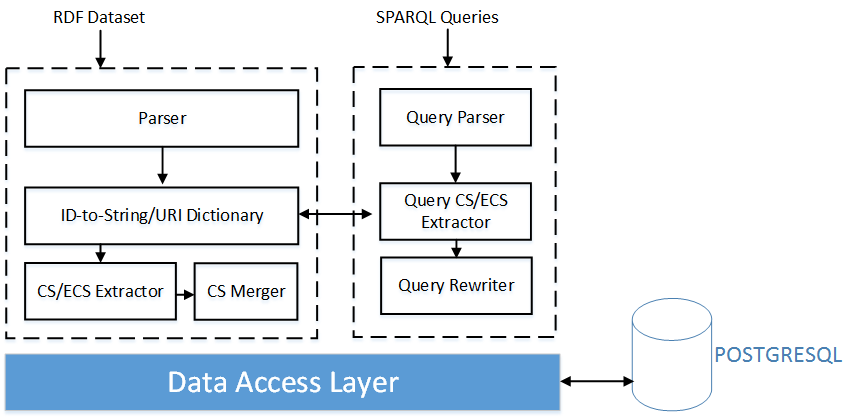}
\caption{Architecture of \textit{raxonDB}.}\label{fig:arch}
\end{figure}

We implemented \textit{raxonDB} as a storage and querying engine that supports hierarchical CS merging, and can be deployed on top of standard RDBMS solutions. Specifically, we used PostgreSQL 9.6, but \textit{raxonDB} can be adapted for other relational databases as well. The architecture of \textit{raxonDB} can be seen in Figure \ref{fig:arch}. 

\textbf{CS Retrieval and Merging. }The processes of retrieving and merging CSs take place during the loading stage of an incoming RDF dataset. CS retrieval is a trivial procedure that requires scanning the whole dataset and storing the unique sets of properties that are emitted from the subject nodes in the incoming triples, and is adopted from our previous work in \cite{meimaris2017extended} where it is described in detail. After retrieving the CSs, the main idea is to compute the inferred CS hierarchy and apply one of the described merging algorithms. Finally, each set of merged CSs is stored in a relational table. In each table, the first column represents the subject identifier, while the rest of the columns represent the union of the property sets of the merged CSs. For multi-valued properties, we use PostgreSQL's array data type in order to avoid duplication of the rows. 


\textbf{Indexing. }We deploy several indexes in \textit{raxonDB}. First off, we index the subject id for each row. We also build foreign-key indexes on object-subject links between rows in different CSs, i.e., when a value of a property in one CS is the subject id of another CS. Next, we use standard B+tree for indexing single-valued property columns, while we use PostgreSQL's GIN indexes, which apply to array datatypes for indexing multi-valued properties. This enables fast access to CS chain queries, i.e., queries that apply successive joins for object-subject relationships. Furthermore, we store these links on the schema level as well, i.e., we keep an index of CS pairs that are linked with at least one object-subject pair of records. These links are called Extended Characteristic Sets (ECSs) and are based on our previous work in \cite{meimaris2017extended}. With the ECS index, we can quickly filter out CSs that are guaranteed not to be related, i.e., no joins exist between them, even if they are individually matched in a chain of query CSs. Other metadata and indexes include the property sets of CSs, and which properties can contain multiple values in the same CS.

\textbf{Query Processing. }Processing SPARQL queries on top of merged CSs entails (i) parsing the queries, (ii) retrieving the query CSs, (iii) identifying the joins between them, and (iv) mapping them to merged tables in the database. Steps (i)-(iii) are inherited from our previous work in \cite{meimaris2017extended}. For (iv), a query CS can match with more than one table in the database. For instance, consider a query containing a chain of three CSs, $q_1 \bowtie q_2 \bowtie q_3$, joined sequentially with object-subject joins. Each query CS $q_i$ matches with all tables whose property sets are supersets of the property set of $q_i$. Thus, each join in the initial query creates a set of \textit{permutations} of table joins that need to be evaluated. For instance, assume that $q_1$ matches with $c_1, c_2$, while $q_2$ matches with $c_3$ and $q_3$ matches with $c_4, c_5$. Furthermore, by looking up the ECS index, we derived that the links $\left[c_1,c_3\right]$, $\left[c_2,c_3\right]$, $\left[c_3,c_4\right]$ and $\left[c_3,c_5\right]$ are all valid, i.e., they correspond to candidate joins in the data. Then, $\left[c_1, c_3, c_4\right]$, $\left[c_1, c_3, c_5\right]$, $\left[c_2, c_3, c_4\right]$ and $\left[c_2, c_3, c_5\right]$ are all valid table permutations that must be processed. Two strategies can be employed here. The first is to join the UNIONs of the matching tables for each $q_i$, and the other is to process each permutation of tables separately and append the results. Given the filtering performed by the ECS indexing approach, where we can pre-filter CSs based on the relationships between them, the UNION would impose significant overhead and eliminate the advantage of ECS indexing. Therefore, we have implemented the second approach, that is, process a separate query for each permutation. Finally, due to the existence of NULL values in the merged tables, we must add explicit IS NOT NULL restrictions for all the properties that are contained in each matched CS and are not part of any other restriction or filter in the original query.

%% file: section4_evaluation.tex
\section{Experimental Evaluation}

We implemented \textit{raxonDB} on top of PostgreSQL\footnote{The code and queries are available in https://github.com/mmeimaris/raxonDB}. We did not extend our previous native RDF implementation of \textit{axonDB} \cite{meimaris2017extended}, because given the underlying relational schema of the CS tables, we decided to rely on a well-established relational engine for both the planning and the execution of queries, instead of re-implementing it. As the focus of this paper is to improve RDF storage and querying efficiency in relational settings, we rely on existing mechanisms within PostgreSQL for I/O operations, physical storage and query planning. In this set of experiments, we report results of implementing $hier\_merge$ with the greedy approximation algorithm, as experimenting with the optimal algorithm failed to finish the merging process even in datasets with small numbers of CSs.

\textbf{Datasets. }For this set of experiments, we used two synthetic datasets, namely \textit{LUBM2000} ($\approx$300m triples), and WatDiv ($\approx$100m triples), as well as two real-world datasets, namely \textit{Geonames} ($\approx$170m triples) and \textit{Reactome} ($\approx$15m triples). LUBM \cite{guo2005lubm} is a customizable generator of synthetic data that describes academic information about universities, departments, faculty, and so on. Similarly, WatDiv\cite{alucc2014diversified} is a customizable generator with more options for the production and distribution of triples to classes. \textit{Reactome}\footnote{http://www.ebi.ac.uk/rdf/services/reactome} is a biological dataset that describes biological pathways, and \textit{Geonames}\footnote{http://www.geonames.org/ontology/documentation.html} is a widely used ontology of geographical entities with varying properties. Geonames maintains a rich graph structure as there is a heavy usage of hierarchical area features on a multitude of levels. 

\textbf{Loading. }In order to assess the effect of hierarchical merging in the loading phase, we performed a series of experiments using all four datasets. For this experiment, we measure the size on disk, the loading time, the final number of merged tables, as well as the number of ECSs (joins between merged tables) and the percentage of triple coverage by CSs included in the merging process, for varying values of the density factor $m \in [0,1]$. The results are summarized in Table \ref{table:loading}. As can be seen, the number of CS, and consequently tables, is greatly reduced with increasing values of $m$. As the number of CSs is reduced, the expected number of joins between CSs is also reduced, which can be seen in the column that measures ECSs. Consequently, the number of tables can be decreased significantly without trading off large amounts of coverage by dense CSs, i.e. large tables with many null values. Loading time tends to be slightly greater as the number of CSs decreases, and thus the number of merges increases, the only exception being WatDiv, where loading time is actually decreased. This is a side-effect of the excessive number of tables ($=5667$) in the simple case  which imposes large overheads for the persistence of the tables on disk and the generation of indexes and statistics for each one.
\begin{figure*}[t]
\def\tabularxcolumn#1{m{#1}}
\begin{tabularx}{\linewidth}{@{}cXX@{}}
\begin{tabular}{ccc}
\subfloat[Execution time (seconds) for LUBM]{\includegraphics[width=0.31\textwidth]{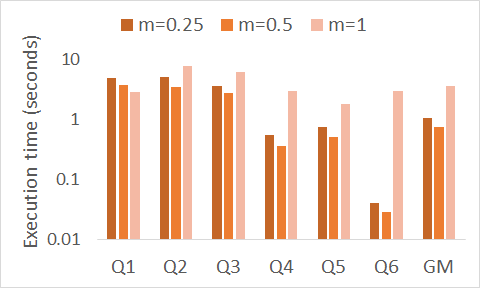}} 
   & \subfloat[Execution time (seconds) for Geonames]{\includegraphics[width=0.3\textwidth]{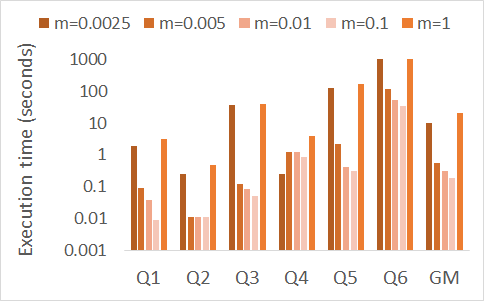}}
& \subfloat[Execution time (seconds) for Reactome]{\includegraphics[width=0.31\textwidth]{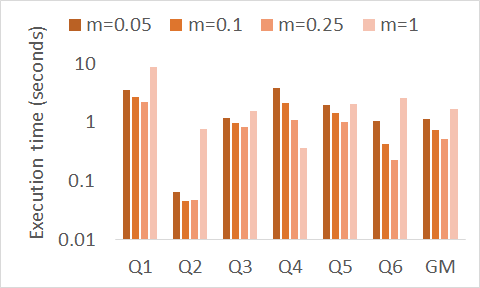}}   
   \\
\end{tabular}
\end{tabularx}
\caption{Query execution times in milliseconds}\label{fig:exp_fig_1}
\end{figure*}

\begin{figure*}[t]
\def\tabularxcolumn#1{m{#1}}
\begin{tabularx}{\linewidth}{@{}cXX@{}}
\begin{tabular}{ccc}
\subfloat[\# of CS permutations for LUBM]{\includegraphics[width=0.31\textwidth]{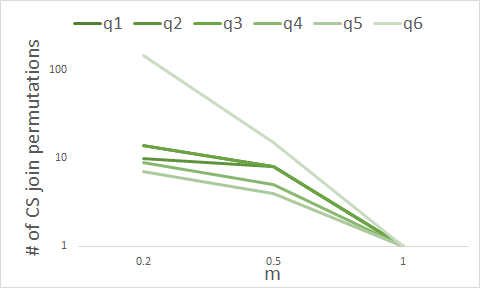}} 
   & \subfloat[\# of CS permutations for Geonames]{\includegraphics[width=0.31\textwidth]{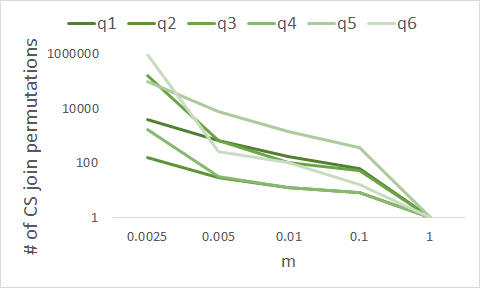}}
& \subfloat[\# of CS permutations for Reactome]{\includegraphics[width=0.31\textwidth]{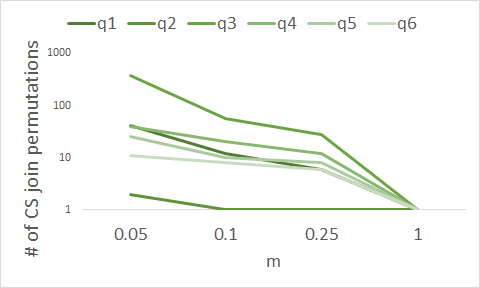}}   
   \\
\end{tabular}
\end{tabularx}
\caption{\# of CS permutations for increasing m}\label{fig:exp_fig_2}
\end{figure*}
\begin{table}
\small
\centering
\caption{Loading experiments for all datasets}
\resizebox{\columnwidth}{!}{
\begin{tabular}{|c|c c c c c |} 
\hline
\textbf{Dataset} & \textbf{Size (MB)} & \textbf{Time} & \textbf{\# Tables (CSs)} & \textbf{\# of ECSs} & \textbf{Dense CS}\\
& & & & & Coverage\\
 \hline
Reactome Simple & 781 & 3min & 112 & 346 & 100\% \\
Reactome (m=0.05) & 675 & 4min & 35 & 252 &  97\% \\
Reactome (m=0.25) & 865 & 4min & 14 & 73 & 77\% \\
 \hline
Geonames Simple & 4991 & 69min & 851 & 12136 & 100\% \\
Geonames (m=0.0025) & 4999 & 70min & 82 & 2455 & 97\% \\
Geonames (m=0.05) & 5093 & 91min & 19 & 76 & 87\% \\
Geonames (m=0.1) & 5104 & 92min & 6 & 28 & 83\% \\
 \hline
LUBM Simple & 591 & 3min & 14 & 68 & 100\% \\
LUBM (m=0.25) & 610 & 3min & 6 & 21 & 90\% \\
LUBM (m=0.5) & 620 & 3min & 3 & 6 & 58\% \\
 \hline
WatDiv Simple & 4910 & 97min & 5667 & 802 & 100\% \\
WatDiv (m=0.01) & 5094 & 75min & 67 & 99 & 77\% \\
WatDiv (m=0.1) & 5250 & 75min & 25 & 23 & 63\% \\
WatDiv (m=0.5) & 5250 & 77min & 16 & 19 & 55\% \\
\hline
\end{tabular}
}
\label{table:loading}
\end{table}

\textbf{Query Performance. }
In order to assess the effect of the density factor parameter $m$ during query processing, we perform a series of experiments on LUBM, Reactome and Geonames. For the workload, we used the sets of queries from \cite{meimaris2017extended}. We employ two metrics, namely \textit{execution time} and \textit{number of table permutations}. The results can be seen in Figures \ref{fig:exp_fig_1} and \ref{fig:exp_fig_2}. As can be seen, hierarchical CS merging can help speed up query performance significantly as long as the dense coverage remains high. For example, in all datasets, query performance degrades dramatically when $m=1$, in which case the merging process cannot find any dense CSs. In this case, all rows are added to one large table, which makes the database only contain one table with many NULL cells. These findings are consistent across all three datasets and require further future work in order to identify the optimal value for $m$. 

\begin{figure*}[t]
\def\tabularxcolumn#1{m{#1}}
\begin{tabularx}{\linewidth}{@{}cXX@{}}
\begin{tabular}{ccc}
\subfloat[Execution time (seconds) for LUBM2000]{\includegraphics[width=0.33\textwidth]{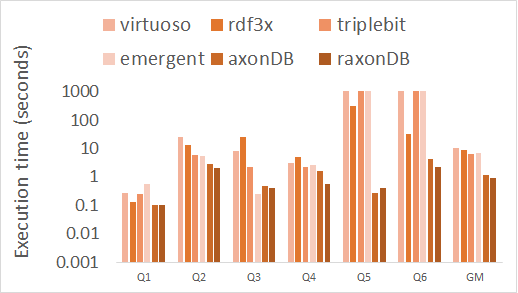}} 
   & \subfloat[Execution time (seconds) for Geonames]{\includegraphics[width=0.33\textwidth]{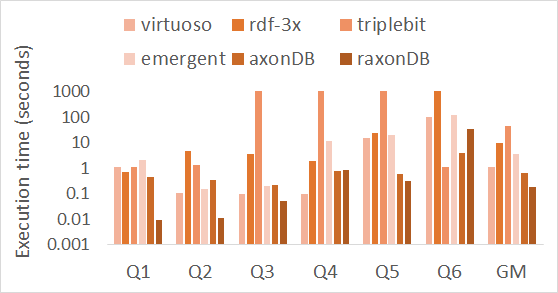}}
& \subfloat[Execution time (seconds) for Reactome]{\includegraphics[width=0.33\textwidth]{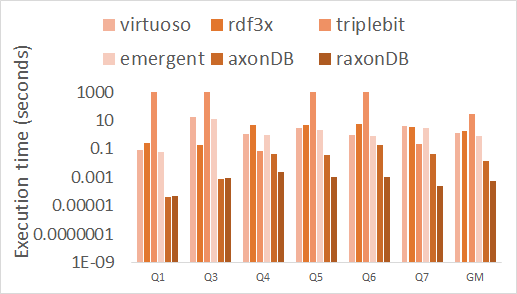}}   
   \\
\end{tabular}
\end{tabularx}
\caption{Query execution times in milliseconds for different RDF engines}\label{fig:comp_with_old}
\end{figure*}

In order to assess the performance of \textit{raxonDB} and establish that no overhead is imposed by the relational backbone, we performed a series of queries on LUBM2000, Geonames and Reactome, assuming the best merging of CSs is employed as captured by $m$ with respect to our previous findings. We also compared the query performance with rdf-3x, Virtuoso 7.1, TripleBit and the emergent schema approach described in \cite{pham2016exploiting}. The results can be seen in Figure \ref{fig:comp_with_old} and indicate that \textit{raxonDB} provides equal or better performance from the original \textit{axonDB} implementation, as well as the rest of the systems, including the emergent schema approach, which is the only direct competitor for merging CSs. Especially for queries with large intermediate results and low selectivity that correspond to a few CSs and ECSs (e.g. LUBM Q5 and Q6, Geonames Q5 and Q6) several of the other approaches fail to answer fast and in some cases time out.

%% file: section5_conclusions.tex
\section{Conclusions and Future Work}
In this paper, we tackled the problem of merging characteristic sets based on their hierarchical relationships. As future work, we will study computation of the optimal value for $m$, taking into consideration workload characteristics as well as a more refined cost model for the ancestral paths. Furthermore, we will study application of these findings in a distributed architecture, in order to further scale the capabilities of \textit{raxonDB}.

%% file: cs_merge.bbl
\begin{thebibliography}{10}

\bibitem{abadi2007scalable}
D.~J. Abadi, A.~Marcus, S.~R. Madden, and K.~Hollenbach.
\newblock Scalable semantic web data management using vertical partitioning.
\newblock In {\em VLDB}, 2007.

\bibitem{alucc2014diversified}
G.~Alu{\c{c}}, O.~Hartig, M.~T. {\"O}zsu, and K.~Daudjee.
\newblock Diversified stress testing of rdf data management systems.
\newblock In {\em ISWC}, 2014.

\bibitem{bornea2013building}
M.~A. Bornea, J.~Dolby, A.~Kementsietsidis, K.~Srinivas, P.~Dantressangle,
  O.~Udrea, and B.~Bhattacharjee.
\newblock Building an efficient rdf store over a relational database.
\newblock In {\em ACM SIGMOD}, 2013.

\bibitem{erling2010virtuoso}
O.~Erling and I.~Mikhailov.
\newblock {\em Virtuoso: {RDF} support in a native RDBMS}.
\newblock Springer, 2010.

\bibitem{gubichev2014exploiting}
A.~Gubichev and T.~Neumann.
\newblock Exploiting the query structure for efficient join ordering in sparql
  queries.
\newblock In {\em EDBT}, 2014.

\bibitem{guo2005lubm}
Y.~Guo, Z.~Pan, and J.~Heflin.
\newblock Lubm: A benchmark for owl knowledge base systems.
\newblock {\em Web Semantics: Science, Services and Agents on the World Wide
  Web}, 3(2):158--182, 2005.

\bibitem{janik2005brahms}
M.~Janik and K.~Kochut.
\newblock Brahms: a workbench rdf store and high performance memory system for
  semantic association discovery.
\newblock In {\em ISWC}, 2005.

\bibitem{meimaris2017extended}
M.~Meimaris, G.~Papastefanatos, N.~Mamoulis, and I.~Anagnostopoulos.
\newblock Extended characteristic sets: Graph indexing for sparql query
  optimization.
\newblock In {\em ICDE}, 2017.

\bibitem{montoya2017odyssey}
G.~Montoya, H.~Skaf-Molli, and K.~Hose.
\newblock The odyssey approach for optimizing federated sparql queries.
\newblock In {\em ISWC}, 2017.

\bibitem{neumann2011characteristic}
T.~Neumann and G.~Moerkotte.
\newblock Characteristic sets: Accurate cardinality estimation for {RDF}
  queries with multiple joins.
\newblock In {\em ICDE}, 2011.

\bibitem{neumann2010RDF}
T.~Neumann and G.~Weikum.
\newblock The {RDF}-3x engine for scalable management of {RDF} data.
\newblock {\em The VLDB Journal}, 19(1):91--113, 2010.

\bibitem{papailiou2014h}
N.~Papailiou, D.~Tsoumakos, I.~Konstantinou, P.~Karras, and N.~Koziris.
\newblock H 2 {RDF}+: an efficient data management system for big {RDF} graphs.
\newblock In {\em ACM SIGMOD}, 2014.

\bibitem{pham2016exploiting}
M.~Pham and P.~Boncz.
\newblock Exploiting emergent schemas to make rdf systems more efficient.
\newblock In {\em ISWC}, 2016.

\bibitem{pham2015deriving}
M.~Pham, L.~Passing, O.~Erling, and P.~Boncz.
\newblock Deriving an emergent relational schema from rdf data.
\newblock In {\em WWW}, 2015.

\bibitem{schatzle2014sempala}
A.~Sch{\"a}tzle, M.~Przyjaciel-Zablocki, A.~Neu, and G.~Lausen.
\newblock Sempala: interactive sparql query processing on hadoop.
\newblock In {\em ISWC}, 2014.

\bibitem{schatzle2016s2rdf}
A.~Sch{\"a}tzle, M.~Przyjaciel-Zablocki, S.~Skilevic, and G.~Lausen.
\newblock S2rdf: Rdf querying with sparql on spark.
\newblock In {\em VLDB}, 2016.

\bibitem{sidirourgos2008column}
L.~Sidirourgos, R.~Goncalves, M.~Kersten, N.~Nes, and S.~Manegold.
\newblock Column-store support for rdf data management: not all swans are
  white.
\newblock In {\em VLDB}, 2008.

\bibitem{weiss2008hexastore}
C.~Weiss, P.~Karras, and A.~Bernstein.
\newblock Hexastore: sextuple indexing for semantic web data management.
\newblock In {\em VLDB}, 2008.

\bibitem{wilkinson2006jena}
K.~Wilkinson.
\newblock Jena property table implementation, 2006.

\bibitem{yuan2013triplebit}
P.~Yuan, P.~Liu, B.~Wu, H.~Jin, W.~Zhang, and L.~Liu.
\newblock Triplebit: a fast and compact system for large scale rdf data.
\newblock In {\em VLDB}, 2013.

\end{thebibliography}
